\documentclass{iau}
\usepackage{graphicx}

\title[Link Z with Atoll] 
{Links Between Z Sources and Atoll Sources}

\author[J. Wang \& H.-K. Chang]   
{Joan Jing Wang $^1$
 \and  Hsiang-Kuang Chang $^{1,2}$}

\affiliation{$^1$ Institute of Astronomy, National Tsing Hua University, \\
Hsinchu 30013, Taiwan \\ email: {\tt jwang@mx.nthu.edu.tw} \\[\affilskip]
$^2$ Department of Physics, National Tsing Hua University, \\ Hsinchu
30013, Taiwan \\email: {\tt hkchang@phys.nthu.edu.tw}}

\pubyear{2012}
\volume{290}  
\jname{Feeding compact objects: Accretion on all scales}
\editors{C.M. Zhang, T. Belloni, M. M\'endez \& S.N. Zhang, eds.}

\begin{document}

\maketitle

\begin{abstract}
It is known that the Z and atoll sources are two typical types of neutron-star sources in low mass X-ray binaries (LMXBs), which present very different Q-$\nu$ relations of lower kHz QPOs. We propose that the Z and atoll sources are two different phases in the evolutionary track of neutron star in LMXBs, instead of two types of distinct sources.
\keywords{LMXBs, Z source, atoll source}
\end{abstract}

\firstsection 
\section{Introduction}

Two typical types of sources, Z and atoll sources, in neutron star (NS) low mass X-ray binaries (LMXBs) describe different tracks in CCDs on
different timescales and display distinct $Q-\nu$ relations of lower kHz QPOs \cite[Hasinger
\& van der Klis (1989)]{Hasinger1989}, \cite[Wang et al. (2012)]{Wang2012}.
However, the $Q-\nu$ relations of lower kHz QPOs of Sco X-1 displays a mildly similar trend to that
of atoll sources \cite[Wang et al. (2012)]{Wang2012}.
Moreover, recent studies
show that the characteristics of both Z and atoll types present in
two sources \cite[Ding et al. (2011)]{Ding2011} Ding et al. 2010 for a review). In addition, a compilation of
RXTE data for three transient atoll sources presents new branch
which connects to the top of the C-shaped (atoll) path and forms a
horizontal track
--- turning C-shape into a "Z" when they go down to very low
luminosity \cite[Gierli$\acute{n}$ski \& Done (2002)]{GD2002}. These phenomena
enlighten the investigation for the links between Z and atoll
sources.

\section{Links Z with Atoll Sources}

Many evidences indicate that there is a connection between Z and atoll sources. We suppose an evolutionary scenario that the LMXBs containing a NS may evolve from
a Cyg-like Z phase at the beginning and transform into the Sco-
like Z phase. Eventually, they enter into atoll phase, going
through a phase characterized by hybrid atoll/Z phase like
the sources GX 13+1 \cite[Hasinger \& van der Klis (1989)]{Hasinger1989}, \cite[Schnerr
et al. (2003)]{Schnerr2003}, \cite[Homan et al. (2004)]{Homan2004}. The variation of accretion rate dominates the whole process.
At the beginning, the matter with high falling velocity releases substantial gravitational energy and forms a radiation
dominant inner disk \cite[Frank et al. (2002)]{Frank2002}.
The accreting material piles up around the NS at high accretion rate (near critical Eddington accretion rate, even
supercritical rate), describing a Z-shape track in CCDs. During
this process, the disk may puff up and become geometrically
thick due to some instabilities \cite[Paczy\'{n}sky \& Wiita (1980)]{Paczynsky1980},
resulting in two effects. Firstly, more and more material pile
up near the polar cap and bury the magnetic field on NS sur-
face \cite[Zhang \& Kojima (2006)]{Zhang2006}, which leads to the decrease of
magnetic pressure. Furthermore, the viscosity and friction
between the accreted matter slow down the falling velocity,
resulting in the decrease of ram pressure and low accretion
rate. This scenario is consistent with the atoll phase. Then
the NB and FB turn into the IS and banana branch from
the Sco-like phase, respectively. Besides, the low accretion
rate results in a long timescale for the trace of atoll patterns
in CCDs, as well as a wide range of luminosity. Because of
the inhomogeneities in the inner accretion disk \cite[Romanova,
Kulkarni \& Lovelace (2007)]{RKL2007}, the friction is different from here
to there. Consequently, the accretion rate may become high
at some position, which generates a horizon pattern connect-
ing to the top of the C-shape for atoll sources, consistent
with the scenario presenting in \cite[Gierli$\acute{n}$ski \& Done (2002)]{GD2002}.

\section{Conclusion}

We suggest that the Z type and atoll type sources are two phases during
different evolutionary phase of NS LMXBs, instead of two distinct sources.


\begin{thebibliography}{}

\bibitem[Ding \etal\ (2011)]{Ding2011} {Ding, G. Q., Zhang, S. N.,
Wang, N., Qu, J. L., \& Yan, S. P.} 2011, \textit{AJ}, 142, 34

\bibitem[Frank \etal\ (2002)]{Frank2002}
Frank J., King A. \& Raine D. 2002, Accretion Power in Astrophysics, (Cambridge, UK: Cambridge
University Press)

\bibitem[Gierli$\acute{n}$ski M. \& Done C. (2002)]{GD2002}
{Gierli$\acute{n}$ski M. \& Done C.} 2002, \textit{MNRAS}, 331, L47

\bibitem[Hasinger \& van der Klis (1989)]{Hasinger1989}
{Hasinger G., \& van der Klis M.} 1989, \textit{A\&A}, 225, 79

\bibitem[Homan \etal\ (2004)]{Homan2004}
{Homan J., Wijnands R., Rupen M. P., Fender R., Hjellming R. M., di
Salvo T., van der Klis M.} 2004, \textit{A\&A}, 418, 255

\bibitem[Paczy$\acute{n}$sky \& Wiita (1980)]{Paczynsky1980}
{Paczy$\acute{n}$sky B. \& Wiita P. J.} 1980, \textit{A\&A}, 88, 23

\bibitem[Romanova, Kulkarni \& Lovelace (2007)]{RKL2007}
Romanova M. M., Kulkarni A. K. \& Lovelace R. V. E. 2007 (astroph/
0711.0418)

\bibitem[Schnerr \etal\ (2003)]{Schnerr2003}
{Schnerr R. S., Reerink T., van der Klis M., Homan J.,
M$\acute{e}$ndez M., Fender R. P., Kuulkers E.} 2003, \textit{A\&A}, 406, 221

\bibitem[Wang \etal\ (2012)]{Wang2012}
{J.Wang, H.K. Chang, C.M. Zhang et al.} 2012,
\textit{Astrophysics and Space Science}, 342, 357

\bibitem[Zhang \& Kojima (2006)]{Zhang2006}
Zhang C. M. \& Kojima Y. 2006, MNRAS, 366, 137




\end{thebibliography}
\end{document}